\documentclass{aa}

\usepackage{amssymb}

\usepackage{graphicx}
\usepackage[varg]{txfonts} 
\usepackage{lineno}
\newcommand{\cotwo}{\mbox{CO$_{2}$}}
\newcommand{\otwo}{\mbox{O$_{2}$}}
\newcommand{\htwoo}{\mbox{H$_{2}$O}}
\newcommand{\htwo}{\mbox{H$_{2}$}}
\newcommand{\lya}{{Lyman-$\alpha$}}
\newcommand{\lyb}{{Lyman-$\beta$}}

\begin{document}

\title{Measurements of the Near-Nucleus Coma of Comet 67P/Churyumov-Gerasimenko with the Alice Far-Ultraviolet Spectrograph on Rosetta} 

\author{Paul D. Feldman
	\inst{1}
	\and
	Michael F. A'Hearn\inst{2}
	\and
	Jean-Loup Bertaux\inst{3}
	\and
	Lori M. Feaga\inst{2}
	\and
	Joel Wm. Parker\inst{4}
	\and
	Eric Schindhelm\inst{4}
	\and
	Andrew J. Steffl\inst{4}
	\and
	S. Alan Stern\inst{4}
	\and
	Harold A. Weaver\inst{5}
	\and
	Holger Sierks\inst{6} 
	\and
	Jean-Baptiste Vincent\inst{6}	
	}

\institute{Johns Hopkins University, Department of Physics and
Astronomy, 3400 N. Charles Street, Baltimore, MD 21218, USA \\ 
	\email{pfeldman@jhu.edu}
	\and
	University of Maryland, Department of Astronomy, College Park, MD 20742, USA
	\and
	LATMOS, CNRS/UVSQ/IPSL, 11 Boulevard d'Alembert, 78280 Guyancourt, France
	\and
	Southwest Research Institute, 
	Department of Space Studies, Suite 300, 
	1050 Walnut Street, 
	Boulder, CO 80302 USA
	\and
	Johns Hopkins University Applied Physics Laboratory, 11100 Johns Hopkins Road,
	Laurel, MD 20723, USA
	\and
	Max Planck Institute for Solar System Research, 37077 G\"ottingen, Germany
	}

   \date{\today}

\abstract
{}
{The Alice far-ultraviolet spectrograph onboard Rosetta is designed to observe emissions from various atomic and molecular species from within the coma of comet 67P/ Churyumov-Gerasimenko and to determine their spatial distribution and evolution with time and heliocentric distance.}
{Following orbit insertion in August 2014, Alice made observations of the inner coma above the limbs of the nucleus of the comet from cometocentric distances varying between 10 and 80 km.  Depending on the position and orientation of the slit relative to the nucleus, emissions of atomic hydrogen and oxygen were initially detected.  These emissions are spatially localized close to the nucleus and spatially variable with a strong enhancement above the comet's neck at northern latitudes.  Weaker emission from atomic carbon and CO were subsequently detected.}  
{Analysis of the relative line intensities suggests photoelectron impact dissociation of \htwoo\ vapor as the source of the observed \ion{H}{i} and \ion{O}{i} emissions.  The electrons are produced by photoionization of \htwoo.  The observed \ion{C}{i} emissions are also attributed to electron impact dissociation, of \cotwo, and their relative brightness to \ion{H}{i} reflects the variation of \cotwo\ to \htwoo\ column abundance in the coma.
}{}

\titlerunning{The Near-Nucleus Coma of Comet 67P/Churyumov-Gerasimenko}
\authorrunning{P. D. Feldman et al.}
   \maketitle


\section{Introduction}
\label{intro}

Over the past 40 years, spectroscopic observations of comets from suborbital and orbital platforms in the vacuum ultraviolet (the spectral range below the atmospheric cut-off at $\sim$3,000~\AA) have yielded a wealth of information about the composition and temporal evolution of cometary comae.  The far-ultraviolet, the region shortward of 2,000~\AA, contains the resonance transitions of the cosmically abundant elements, as well as electronic transitions of simple molecules such as CO and \htwo.  The principal excitation mechanism in the ultraviolet is resonance fluorescence of solar radiation but the relative paucity of solar ultraviolet photons has limited observations from orbiting observatories such as the International Ultraviolet Explorer, the  Hubble Space Telescope (HST) \citep{Lupu:2007,Weaver:2011}, and the Far Ultraviolet Spectroscopic Explorer \citep{Feldman:2002,Weaver:2002} to moderately active comets within $\sim$1.5~AU of the Sun.  The highest spatial resolution at a comet that has been obtained is with the Space Telescope Imaging Spectrograph (STIS) on HST, but is typically limited to tens of kilometers for comets passing closest to the Earth.

The Alice far-ultraviolet spectrograph on Rosetta makes possible observations on scales of tens to hundreds of meters near the nucleus of the comet, and we report on such observations made beginning in August 2014 when the spacecraft entered an orbit $\sim$100~km about the nucleus. Prior to the deployment of the lander, Philae, on November 12, 2014, the spacecraft attained a bound orbit of $\sim$10~km.  The recorded spectra are quite different, and quite unexpected, compared to previously observed coma spectra.  They present a unique opportunity to probe the coma-nucleus interaction region that is not accessible to the in situ instruments on Rosetta or to observations from Earth orbit.  In this initial report we present a sample of the spectra obtained under three different viewing scenarios and discuss our interpretation in terms of photoelectron dissociative excitation of \htwoo.  Our results are supported by concurrent data obtained by other instruments on Rosetta.

\section{Instrument Description}

Alice is a lightweight, low-power, imaging spectrograph optimized for in situ far-ultraviolet (FUV) spectroscopy of comet 67P. It is designed to obtain spatially-resolved spectra in the 700-2050~\AA\ spectral band with a spectral resolution between 8 and 12~\AA\ for extended sources that fill its field-of-view.  The slit is in the shape of a dog bone, 5.5\degr\ long, with a width of 0.05\degr\ in the central 2.0\degr, while the ends are 0.10\degr\ wide.  Each spatial pixel or row along the slit is 0.30\degr\ long.  Alice employs an off-axis telescope feeding a 0.15-m normal incidence Rowland circle spectrograph with a concave holographic reflection grating. The imaging microchannel plate detector utilizes dual solar-blind opaque photocathodes (KBr and CsI) and employs a two-dimensional delay-line readout.  Details of the instrument have been given by \citet{Stern:2007}.

\section{Observations}

\subsection{September Observations}

\begin{figure*}[ht]
\begin{center}
\includegraphics*[width=0.38\textwidth,angle=0.]{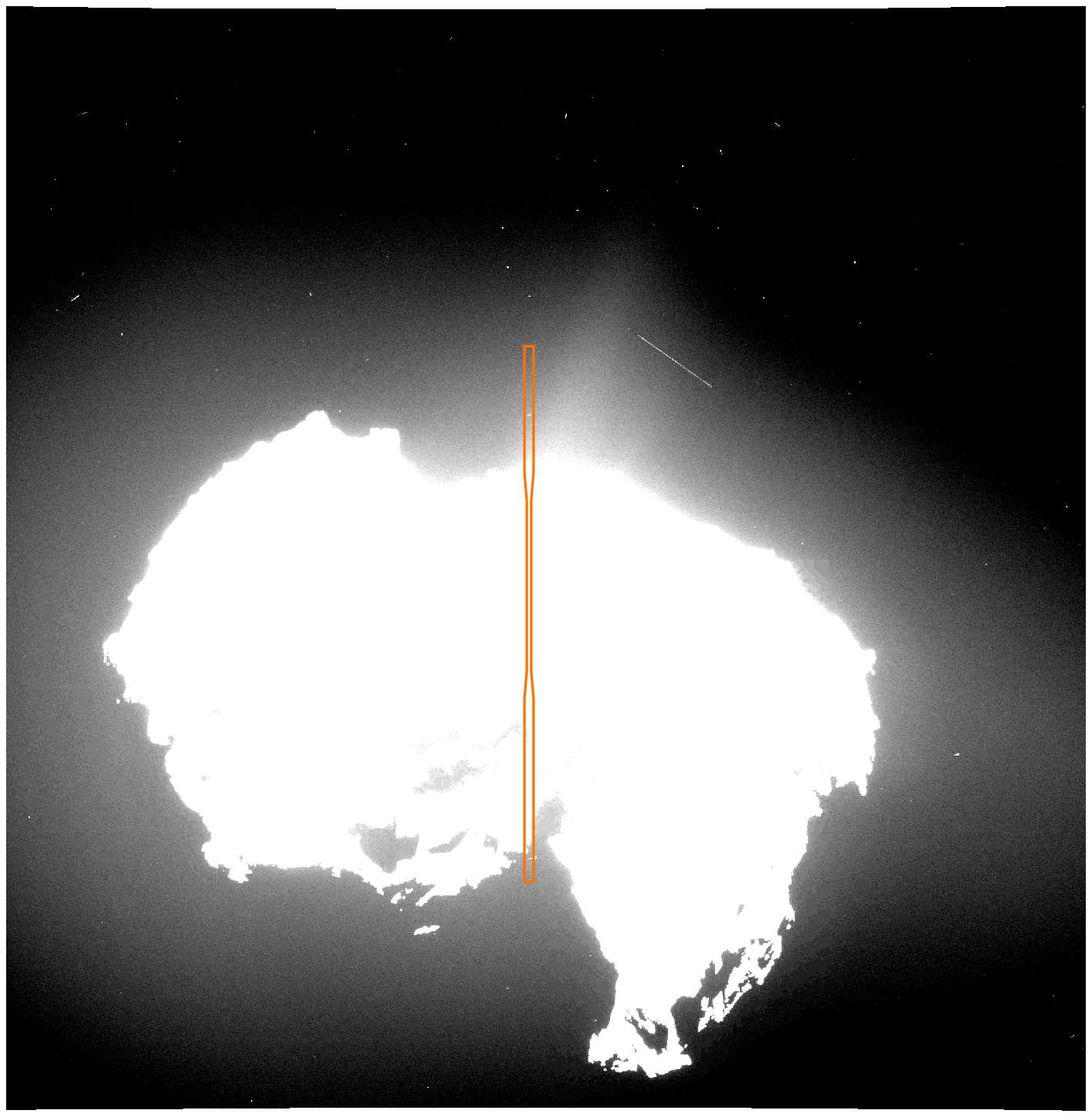}
\includegraphics*[width=0.60\textwidth,angle=0.]{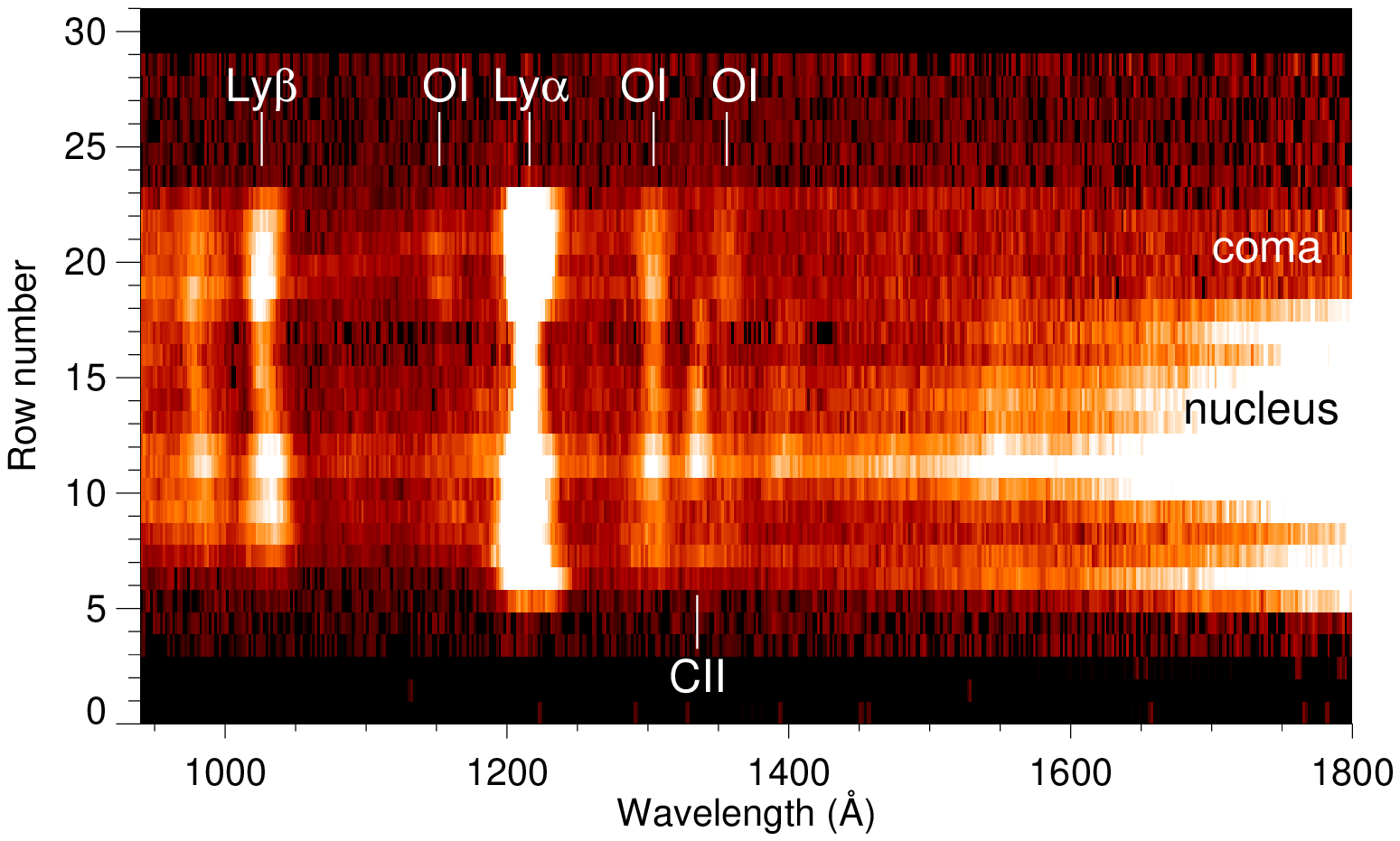}
\caption[]{Left: OSIRIS Wide Angle Camera image taken at UT 01:10 September 21 stretched to show jets above the comet's neck.  The field of view is 12\degr\ $\times$ 12\degr\ which projects to 5.85 $\times$ 5.85~km$^2$ at the center of the nucleus.  The position of the Alice slit is superimposed on the images.  The Sun is towards the top.  Right: A 30-minute spectral image starting at UT 00:07 September 21.  The comet was 3.33~AU from the Sun and the distance to the center of the nucleus was 27.6~km.  The solar phase angle was 62.3\degr.  One spatial row of the spectral image projects to 145 m.  Atomic H and O features seen in the coma (rows 18--23) are identified.  Note the clear separation of \ion{O}{i} $\lambda$1356 in the coma from \ion{C}{ii} $\lambda$1335 in the solar spectrum reflected from the nucleus.  Credits: ESA/Rosetta/MPS for OSIRIS Team MPS/UPD/LAM/IAA/SSO/INTA/UPM/DASP/IDA  \label{jet1} }
\end{center}
\end{figure*}

\begin{figure*}[ht]
\begin{center}
\includegraphics*[width=0.38\textwidth,angle=0.]{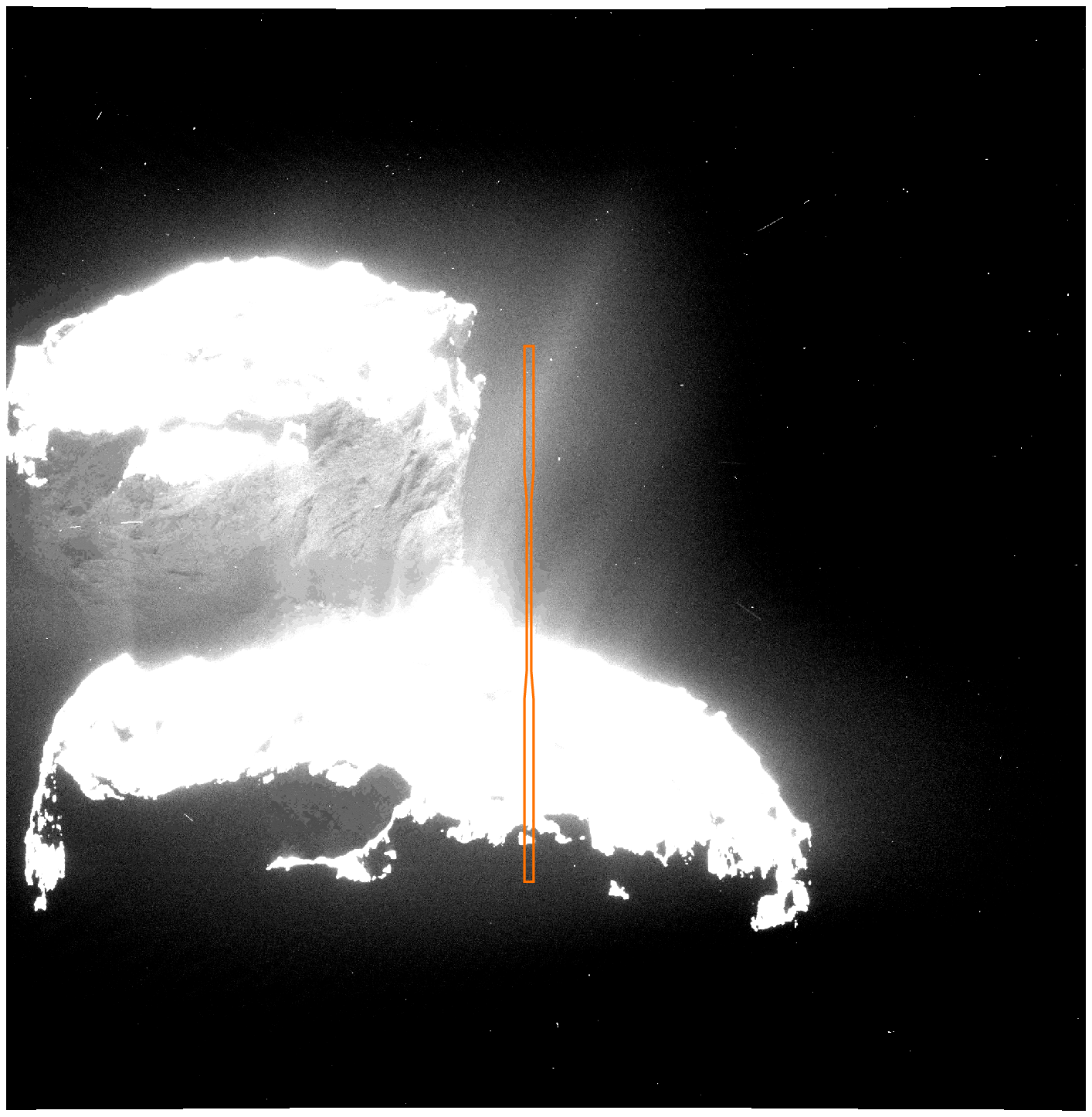}
\includegraphics*[width=0.60\textwidth,angle=0.]{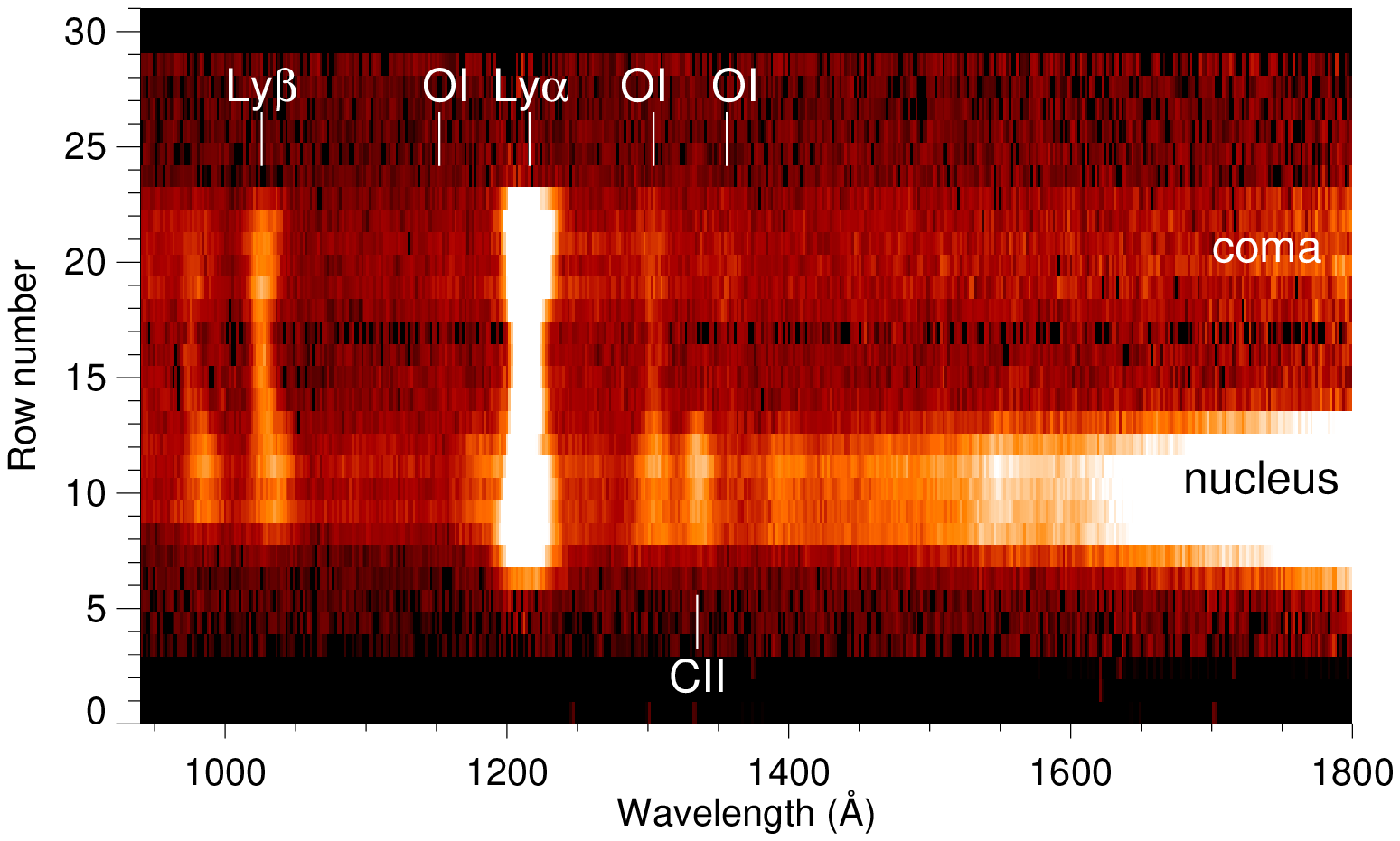}
\caption[]{Left: Same as Fig.~\ref{jet1} for UT 01:28 September 23.  Right: A 50-minute spectral image starting at UT 00:54 September 23.  The comet was 3.32~AU from the Sun and the distance to the center of the nucleus was 28.5~km.  The solar phase angle was 76.1\degr.  One spatial row of the spectral image projects to 150 m.  \label{jet2} }
\end{center}
\end{figure*}

In mid-August 2014, following orbit insertion around 67P on August 6, Rosetta maintained a distance of between 60 and 100~km from the center of the nucleus.  During this period, when the remote sensing instruments were usually pointed at the nadir, coma above the comet's limbs could be detected in the pixels at the wide ends of the Alice slit.  As the distance to the comet was reduced for insertion into a bound orbit at $\sim$30~km, the comet overfilled the slit and coma observations were possible only when the spacecraft was pointed away from nadir towards one of the limbs.  Owing to the irregular shape of the nucleus and residual uncertainties in the spacecraft's trajectory, the exact pointing geometry had to be reconstructed from images taken either by the navigation camera or the OSIRIS Wide Angle Camera \citep[WAC,][]{Sierks:2015}.  Examples of the latter from September 21 and 23 are shown in the left panels of Figures~\ref{jet1} and \ref{jet2}, respectively, in which the brightness scale has been stretched to show the dust jets emanating from the ``neck'' region of the comet.  Both images were taken at a distance of $\sim$28~km from the nucleus.  The projection of the Alice slit on the comet is shown.  Nearly simultaneous spectral images, shown in the right panels of the figures, reveal enhanced emission from atomic hydrogen and oxygen in the region of the slit coincident with the dust jet.  Note that these images represents ``snapshots'' in time.  Our spectra are usually accumulated in 10-minute integrations, a number of which are then co-added to enhance the signal-to-noise ratio of the spectra.  Since the comet is rotating with a 12.4~hour period and the spacecraft is slowly moving, the resulting spectrum does not map exactly onto the image as implied by the figures.

Coma emission is usually quite weak compared with the reflected solar spectrum from the nucleus.  There is also a background of interplanetary \lya\ and \lyb\ emission, the strong \lya\ contributing to a background of grating scattered light over the entire spectral range that can mask weak emission features.  The coma can be identified by a spectrum that contains several features that are weak in the solar spectrum and do not appear in the reflected light from the nucleus.  Both of the spectral images show an unusually bright coma in the high number rows of the Alice detector.  The reflected solar radiation from the nucleus, in the low number rows, is somewhat attenuated by the large solar phase angles of 62.3\degr\ and 76.1\degr, respectively \citep{Feaga:2014}.  Note the presence of two atomic oxygen lines in the coma that do not appear in the spectrum of the nucleus.

\begin{figure*}[ht]
\begin{center}
\includegraphics*[width=0.48\textwidth,angle=0.]{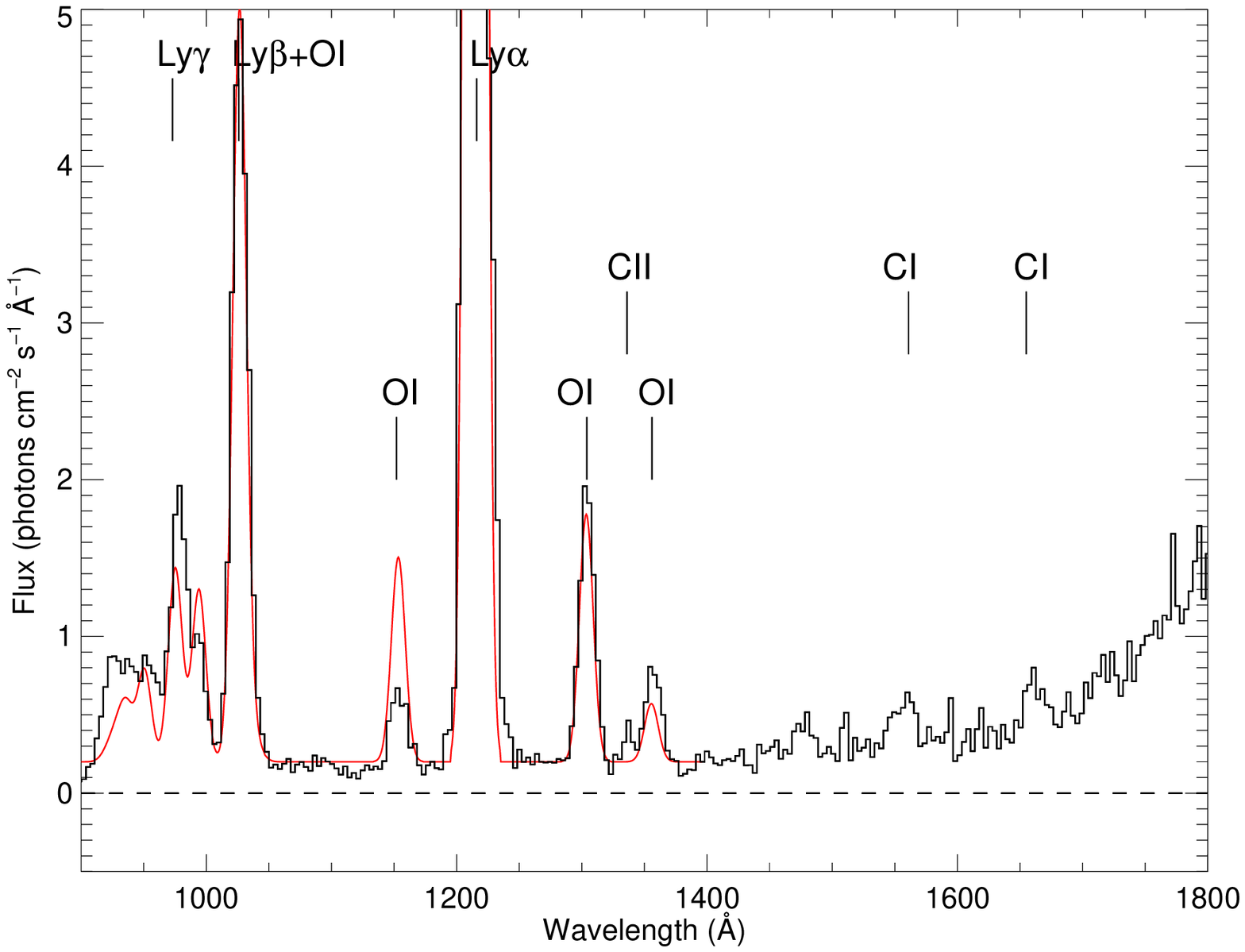}
\includegraphics*[width=0.48\textwidth,angle=0.]{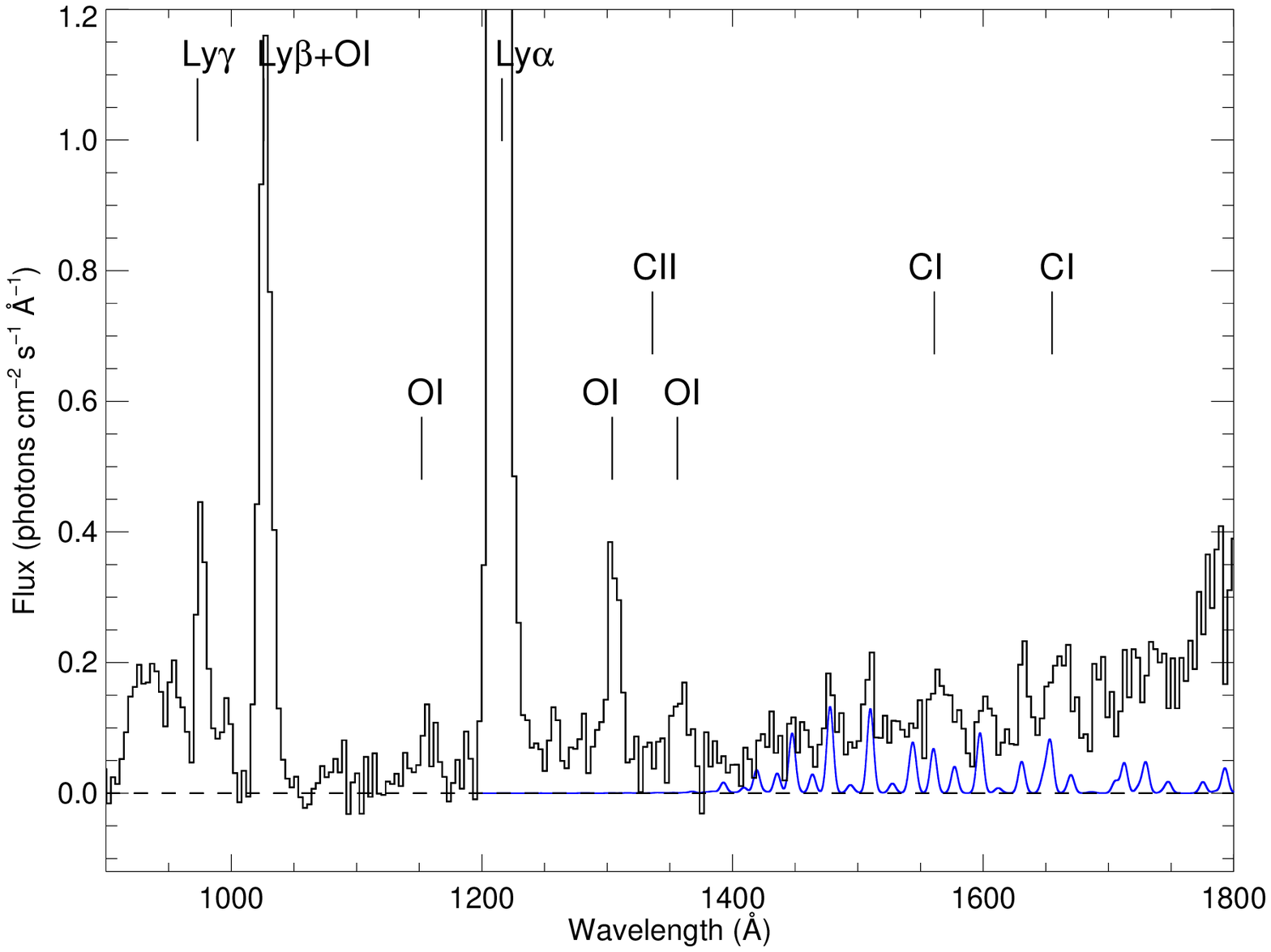}
\caption[]{Coma spectra. Left: a 30-minute exposure beginning at UT 00:07 September 21.  The spectrum is the sum of rows 18--21 and the flux is given for a slit area of 0.1\degr\ $\times$ 1.2\degr\ (48 $\times$ 580~m$^2$ projected on the nucleus).  The red line is a synthetic spectrum of 200~eV electron impact on \htwoo\ (see text).  Right: a 50-minute exposure beginning at UT 00:54 September 23.  The spectrum is the sum of rows 14--17 and the flux is given for a slit area of 0.05\degr\ $\times$ 1.2\degr\ (24 $\times$ 580~m$^2$ projected on the nucleus).  A synthetic spectrum of solar induced fluorescence for a CO column density of $5 \times 10^{13}$~cm$^{-2}$ is shown in blue.  \label{sep_spec} }
\end{center}
\end{figure*}

Spectra of the coma derived from the spectral images of Figures~\ref{jet1} and \ref{jet2} are shown in Figure~\ref{sep_spec}.  The features identified in the figure include the first three members of the \ion{H}{i} Lyman series, \ion{O}{i} multiplets at 1152, 1304, and 1356~\AA, and weak multiplets of \ion{C}{i} at 1561 and 1657~\AA.  In addition, Lyman-$\delta$ and blended higher members of the Lyman series are seen at 950~\AA\ and down to the series limit at 912~\AA.  \ion{O}{i} $\lambda$989 is also partially resolved.  CO Fourth Positive band emission is seen in the spectrum from September 23.  The surprise is \ion{O}{i}] $\lambda$1356, which is a forbidden intercombination multiplet ($^5S^o - ^3P$) that is rarely seen in coma spectra. In the very bright comet C/1995 O1 (Hale-Bopp), its brightness was 42 times smaller than \ion{O}{i} $\lambda$1304 \citep{McPhate:1999}.  The spectra also show a quasi-continuum that we attribute to scattering from residual gas in the vicinity of the spacecraft as its brightness appears correlated with the time following an orbital correction maneuver (thruster firing).

The relative intensities of the observed \ion{H}{i} and \ion{O}{i} multiplets are characteristic of electron dissociative excitation of \htwoo\ \citep{Makarov:2004,Itikawa:2005,McConkey:2008}.  We can exclude prompt photodissociation of \htwoo\ based on both the low photodissociation rate \citep{Wu:1993} and the selection rules for \htwoo\ absorption \citep{Wu:1988}.  Electron impact on \cotwo\ is the likely source of the weak \ion{C}{i} $\lambda\lambda$1561 and 1657 multiplets and the very weak \ion{C}{ii} $\lambda$1335 multiplet \citep{Ajello:1971b,Ajello:1971a,Mumma:1972}.  The source of the electrons is photoionization of \htwoo\ and the resultant photoelectrons have peak energies in the range of 25 to 50 eV, although more energetic electrons have also been detected by the Ion and Electron Sensor (IES) instrument on Rosetta (J. L. Burch, private communication).  Using the cross sections of \citet{Makarov:2004} which are for an incident electron energy of 200 eV, the only energy for which all of the spectral features have measured values, we can calculate a synthetic spectrum for comparison with the data.  This is shown overplotted in red on the spectrum in Figure~\ref{sep_spec}, normalized to the \lyb\ flux, which also gives an excellent fit to \ion{O}{i} $\lambda$1304.  We use \lyb\ for normalization rather than \lya\ because of the difficulty in subtracting the interplanetary background, which is $\sim$300 times brighter at \lya\ than at \lyb.  The other features are fit less well but the differences can probably be attributed to the uncertainties in the electron energy spectrum and cross sections at the lower energies where the photoelectron flux peaks.  In the case of \ion{O}{i}] $\lambda$1356, \citet{Makarov:2004} noted that their laboratory cross section is only an estimate because of the 180~$\mu$s lifetime of the $^5S^o$~state, which leads to a loss of emitters in the interaction region sampled by the experiment.

\subsection{October Off-nadir Limb Stares}

\begin{figure*}[ht]
\begin{center}
\includegraphics*[width=1.0\textwidth,angle=0.]{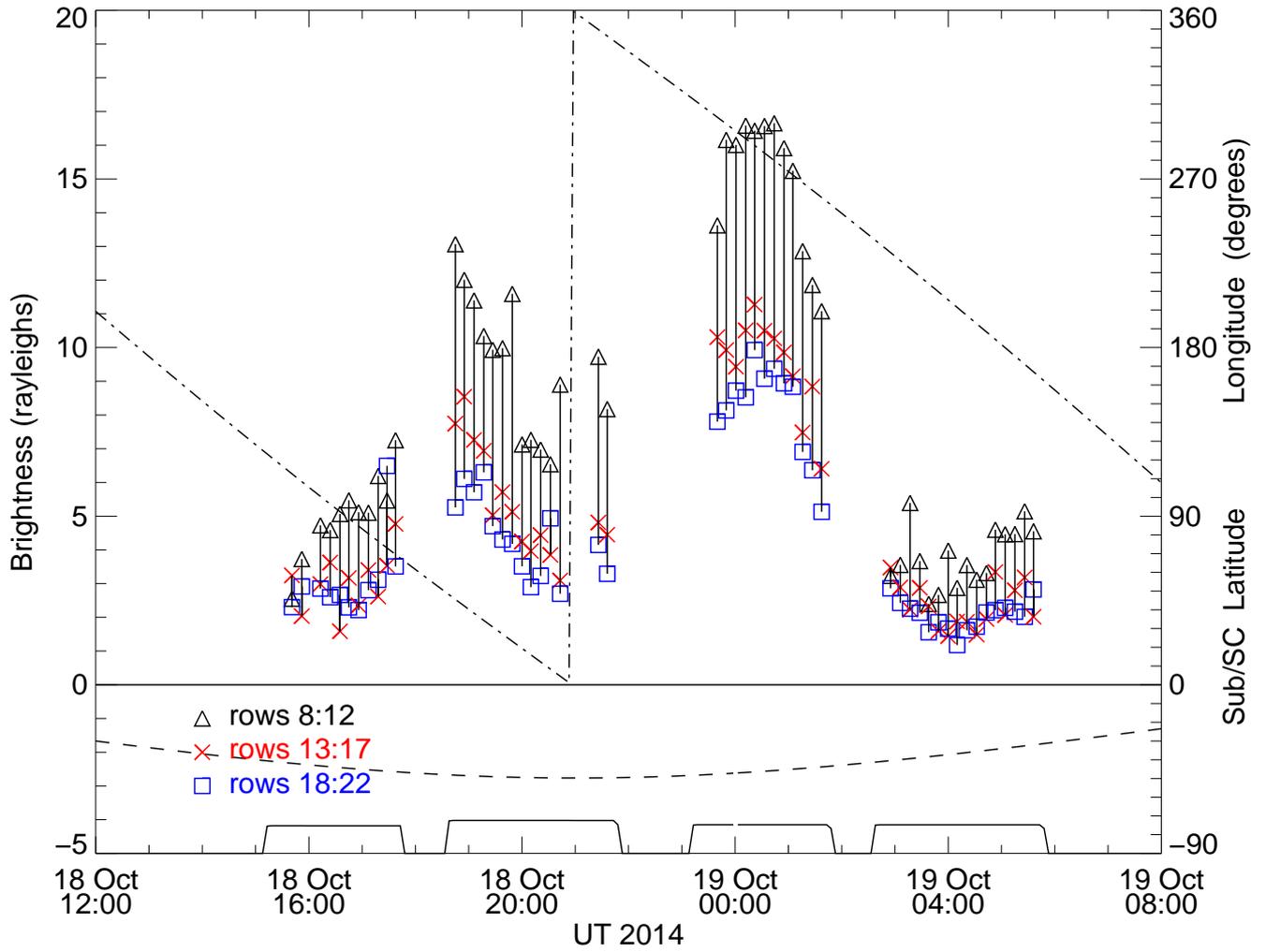}
\caption[]{\ion{H}{i} Lyman-$\beta$ brightness averaged over three regions of the Alice slit (black - wide bottom, red - narrow center, blue - top wide) as a function of time during the 15\degr\ off-nadir limb stares on October 18--19, 2014.  The distance to the comet center was 9.9 km and the heliocentric distance was 3.19 AU.  \label{off1} }
\end{center}
\end{figure*}

\begin{figure*}[ht]
\begin{center}
\includegraphics*[width=1.0\textwidth,angle=0.]{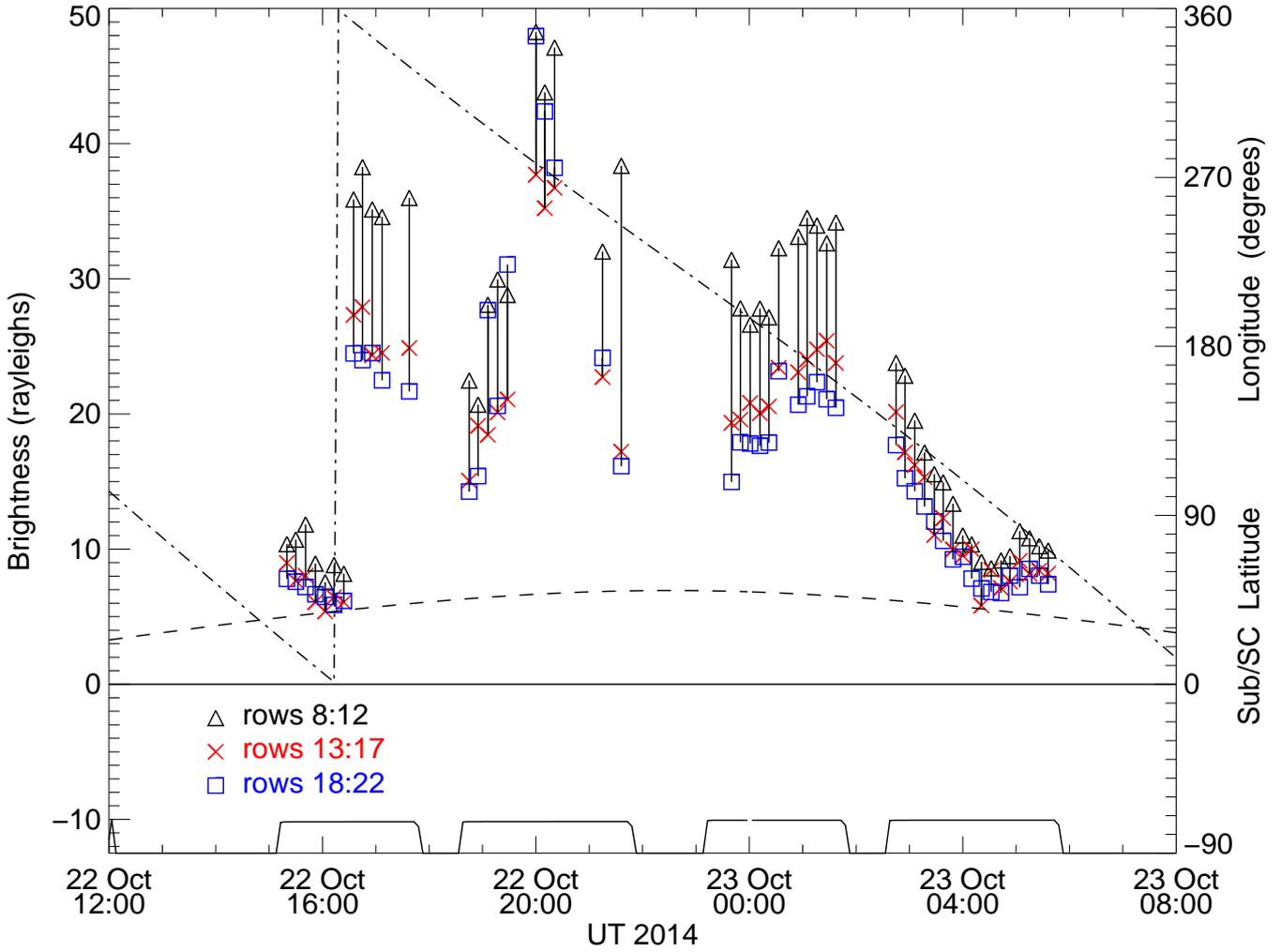}
\caption[]{Same as Fig.~\ref{off1} for the 17\degr\ off-nadir limb stares on October 22--23, 2014.  The distance to the comet center was 9.7 km and the heliocentric distance was 3.13 AU.  \label{off2} }
\end{center}
\end{figure*}

From the 10 km orbit in mid-October 2014 Alice observed emissions from atomic H and O above the comet's limb.  In Figures~\ref{off1} and \ref{off2}, the brightness of the \ion{H}{i} Lyman-$\beta$ line from two off-nadir sequences is used as a surrogate for \htwoo\ abundance along the line-of-sight.  The times of the off-nadir limb stares are indicated at the bottom of each plot.  The vertical bars represent a single 10-minute spectral histogram.  They are {\it not} error bars but represent the average brightness in the wide bottom (black), narrow center (red) and wide top (blue) rows of the Alice slit, and indicate decreasing brightness with distance above the limb (black closest to limb).  The projected separation between the top and bottom points is $\sim$500~m, which represents the apparent scale length of the emissions above the limb.  The sub-spacecraft latitude (dash) and longitude (dot-dash) are also shown in the figures.  Note that the Alice slit appears to cross the boundary of a jet at UT 16:00 on October 22.  The atomic emissions exhibit brightness variations with sub-spacecraft longitude and latitude similar to abundance variations observed by other Rosetta orbiter instruments (ROSINA \citep{Hassig:2015}, MIRO \citep{Gulkis:2015}) with a pronounced enhancement at northern latitudes (Fig.~\ref{off2}) relative to southern latitudes (Fig.~\ref{off1}).

\begin{figure*}[ht]
\begin{center}
\includegraphics*[width=0.48\textwidth,angle=0.]{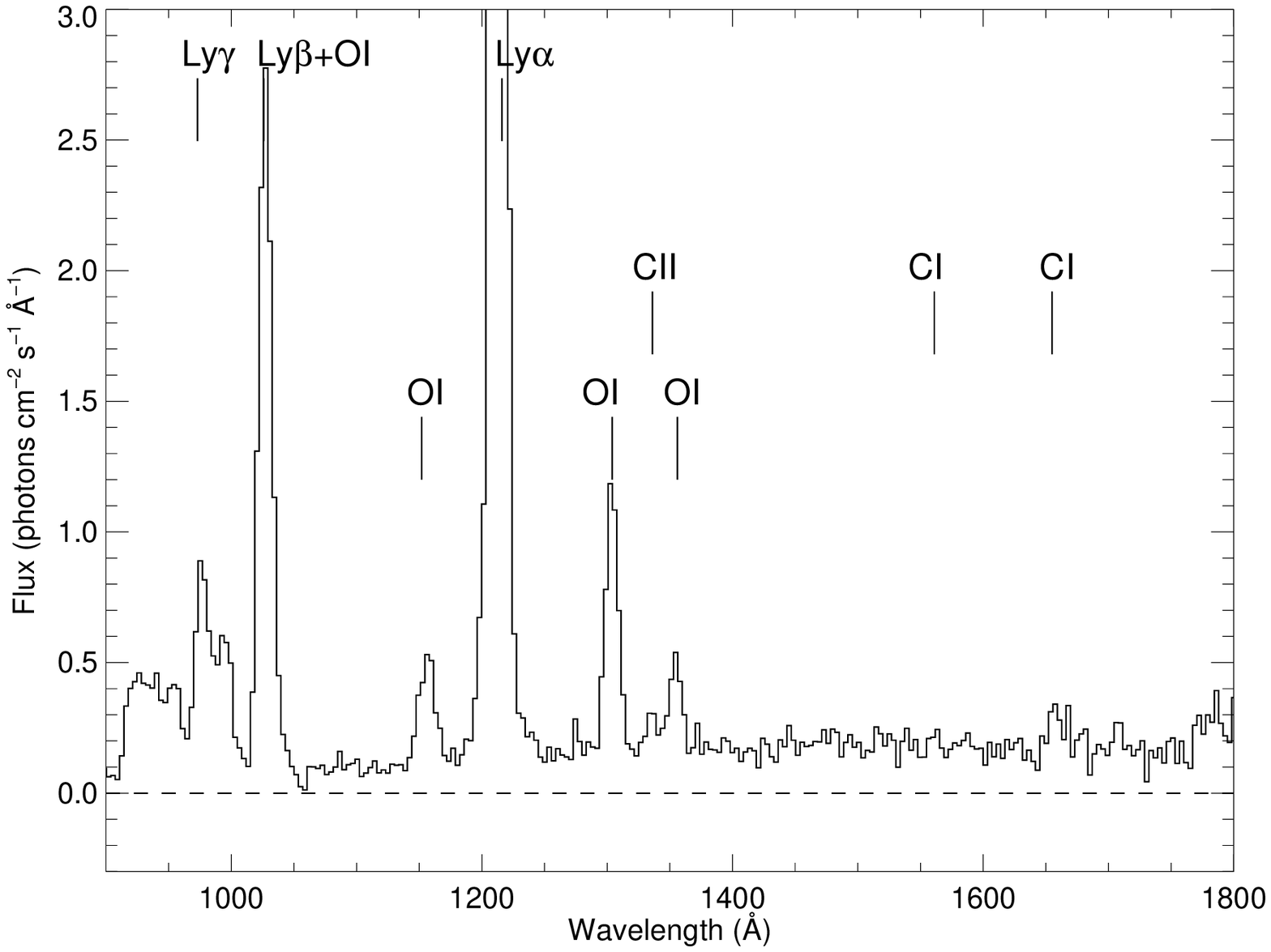}
\includegraphics*[width=0.48\textwidth,angle=0.]{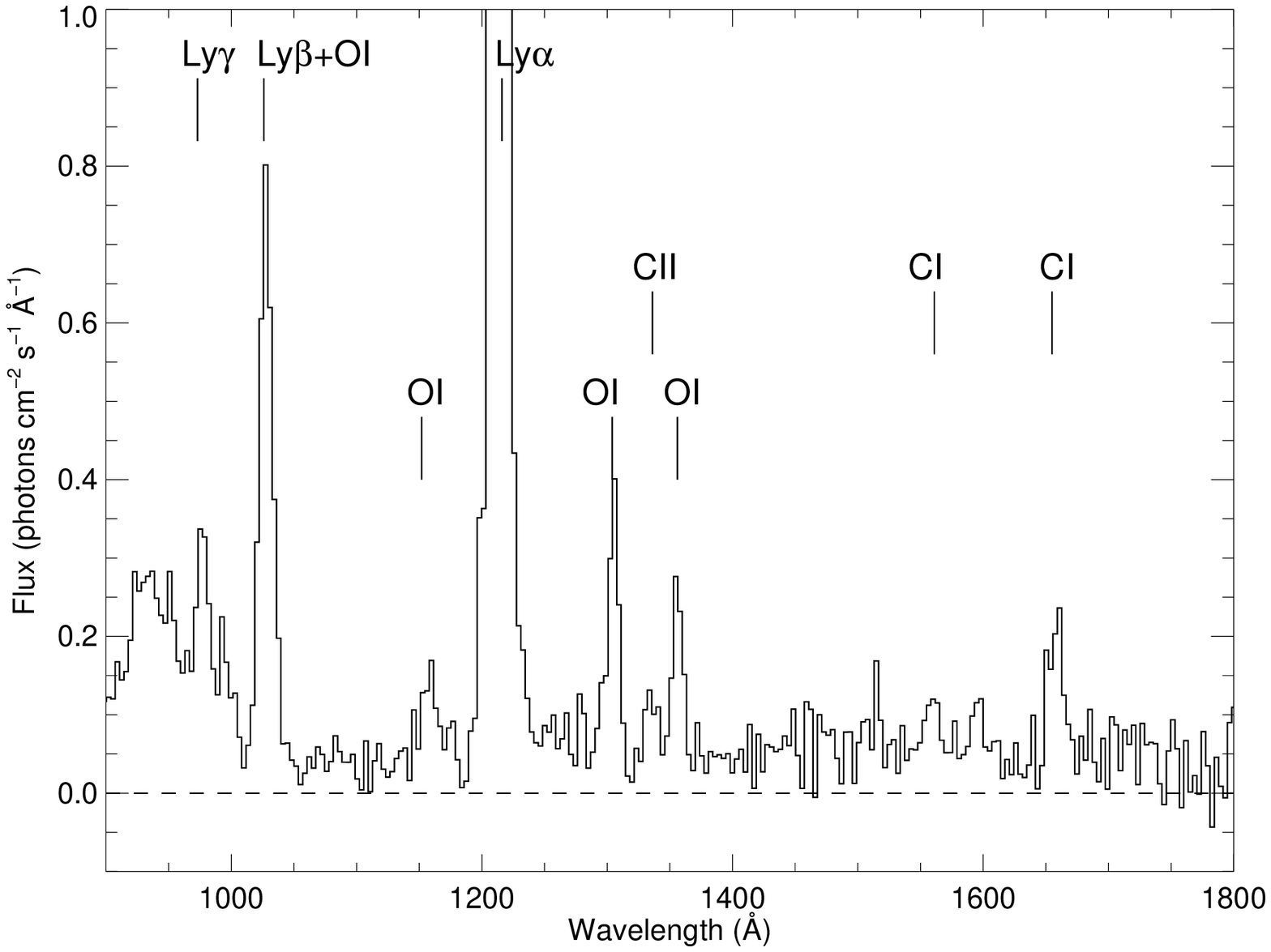}
\caption[]{Spectra from the narrow center of the Alice slit for two periods of Fig.~\ref{off2}, UT 16:30 on October 22 and UT 04:05 on October 23.  The spectrum on the left was taken about 4 hours after the end of an orbit correction maneuver and shows an elevated quasi-continuum, possibly from residual thruster gas in the vicinity of the spacecraft.  \label{stare} }
\end{center}
\end{figure*}

To examine how the spectrum varies with time and spacecraft position, we have selected four periods during the off-nadir stares on October 22-23 (Fig.~\ref{off2}) and co-added multiple 10-minute histograms to enhance the signal-to-noise ratio of the spectra from the narrow center of the Alice slit.  The brightnesses of the brightest atomic emissions are tabulated in Table~\ref{oct_table} and the spectra from the first and last periods are shown in Figure~\ref{stare}.  Note the significant change in the relative brightness of the \ion{C}{i} multiplets to the \ion{H}{i} and \ion{O}{i} emissions between the first and last of these spectra.  As noted above, the \ion{C}{i} (and \ion{C}{ii}) multiplets likely arise from electron impact excitation of either CO or \cotwo.  The relative intensities 1657:1561:1335 $\approx$2:1:1 are more consistent with dissociative excitation of \cotwo\ \citep{Ajello:1971a,Mumma:1972} than with CO \citep{Ajello:1971b} even though the laboratory data on CO are incomplete because of the dominance of CO Fourth Positive emission in the experiment.  Nevertheless, the data of \citet{Ajello:1971b} and of \citet{Aarts:1970} suggest that if electron impact on CO was significant then \ion{C}{ii} $\lambda$1335 would be considerably brighter than \ion{C}{i} $\lambda$1657.

\begin{table*}[ht]
\caption{October 2014 Off-nadir Observations. }
\label{oct_table}
\centering
\begin{tabular}{@{}lccccc@{}}
\hline\hline
Observation Start & Exposure & \multicolumn{4}{c}{Brightness (rayleighs)}   \\
(UT) & Time (s)  & \ion{H}{i} Lyman-$\beta$ & \ion{O}{i} $\lambda$1304 & \ion{O}{i}] $\lambda$1356 & \ion{C}{i} $\lambda$1657 \\
\hline
2014-10-22 16:30:14	& 2419 & $24.0 \pm 0.42$ & $8.68 \pm 0.35$ & $2.91 \pm 	0.35$ & $1.76 \pm 0.74$ \\
2014-10-23 00:07:15	& 5368 & $21.2 \pm 0.37$ & $7.20 \pm 0.32$ & $3.24 \pm 	0.32$ & $2.04 \pm 0.57$ \\
2014-10-23 02:40:01	& 5443 & $12.0 \pm 0.31$ & $4.29 \pm 0.24$ & $1.76 \pm 	0.24$ & $1.23 \pm 0.32$ \\
2014-10-23 04:05:58	& 5443 & $7.13 \pm 0.25$ & $2.39 \pm 0.20$ & $1.48 \pm 	0.20$ & $1.73 \pm 0.38$ \\
\hline
\end{tabular}
\end{table*}

Using cross sections at 100 eV, we can derive the relative \cotwo/\htwoo\ abundance from the brightness ratio of \ion{C}{i} $\lambda$1657 to Lyman-$\beta$, as given for the data of Table~\ref{oct_table} in Table~\ref{oct_ratios}.    The trend of increasing relative \cotwo\ abundance is consistent with the range of relative abundances derived from infrared spectra obtained by the VIRTIS-H instrument during the same time period \citep{Bockelee:2014}.  Note that the data suggest a relatively constant \cotwo\ column during this period while the \htwoo\ column decreased by over a factor of three, consistent with the trend reported by ROSINA \citep{Hassig:2015}.  A detailed study of the abundance ratio as a function of position above the limb will be presented in a subsequent paper.

\begin{table*}[ht]
\caption{Brightness Ratios and Derived \cotwo/\htwoo\ Column Abundance Ratios. }
\label{oct_ratios}
\centering
\begin{tabular}{@{}lcccc@{}}
\hline\hline
Observation Start & \ion{O}{i} $\lambda$1304/ & \ion{O}{i}] $\lambda$1356/ & \ion{C}{i} $\lambda$1657/ & [\cotwo]/  \\
(UT) & \ion{H}{i} Lyman-$\beta$ & \ion{O}{i} $\lambda$1304 & \ion{H}{i} Lyman-$\beta$ & [\htwoo] \\
\hline
2014-10-22 16:30:14	& $0.362 \pm 0.044$ & $0.335 \pm 0.043$ & $0.073 \pm 0.031$ & $0.045 \pm 0.019$ \\
2014-10-23 00:07:15	& $0.340 \pm 0.034$ & $0.450 \pm 0.049$ & $0.096 \pm 0.027$ & $0.059 \pm 0.017$ \\
2014-10-23 02:40:01	& $0.359 \pm 0.050$ & $0.410 \pm 0.061$ & $0.103 \pm 0.027$ & $0.064 \pm 0.016$ \\
2014-10-23 04:05:58	& $0.336 \pm 0.048$ & $0.618 \pm 0.100$ & $0.242 \pm 0.054$ & $0.150 \pm 0.033$ \\
\hline
\end{tabular}
\end{table*}

Table~\ref{oct_ratios} also shows an interesting trend in the ratio of \ion{O}{i}] $\lambda$1356 to \ion{O}{i} $\lambda$1304.  \citet{Makarov:2004} give a value of 0.23 for this ratio from electron impact on \htwoo\ at 100~eV together with their caveat about the approximate nature of their derived value of the \ion{O}{i}] $\lambda$1356 cross section.  Most of the early Alice observations of coma spectra show a ratio of between 0.3 and 0.35, somewhat higher than the laboratory value but compatible with the caveat of \citeauthor{Makarov:2004}  However, as can be seen in Table~\ref{oct_ratios}, the spectra from three of the four observations reflect a higher brightness ratio and follow the trend of increasing \cotwo/\htwoo\ column abundance.  This is suggestive of electron impact excitation of either \cotwo\ and/or \otwo, both of which have \ion{O}{i}] $\lambda$1356 cross sections much larger than that of \htwoo\ and also larger than the corresponding cross section for excitation of \ion{O}{i} $\lambda$1304.

For \cotwo, \citet{Wells:1972a} reported a cross section for electron impact excitation of \ion{O}{i}] $\lambda$1356 at 100~eV $\sim$100 times higher than that for \htwoo.  \citet{Wells:1972b} subsequently revised the number down by a factor of 3 based on a laboratory measurement of the O($^5{\rm S}^o$) lifetime.  \citet{McConkey:2008} note the conflicting reports of the measured \ion{O}{i} $\lambda$1304 cross section at 100~eV but nevertheless the cross section for \ion{O}{i}] $\lambda$1356 is larger by a factor of 2 or 3 depending on whose cross section is adopted, and electron excitation of \cotwo\ can account for the increase in the intensity ratio derived in Table~\ref{oct_ratios}.  Regrettably, the laboratory data currently available do not allow for a more quantitative analysis.  We also note that the electron impact cross sections for \otwo\ to produce the atomic oxygen emissions are very similar to those for \cotwo, with the \ion{O}{i}] $\lambda$1356 to \ion{O}{i} $\lambda$1304 ratio well known to be 2.2 \citep{Kanik:2003}, so that \otwo, recently reported in ROSINA mass spectra (K. Altwegg, private communication), could also be contributing to the intensity ratio increase observed by Alice.

\subsection{Observations Against and Above the Dark Neck - November 29}

\begin{figure*}[ht]
\begin{center}
\includegraphics*[width=0.37\textwidth,angle=0.]{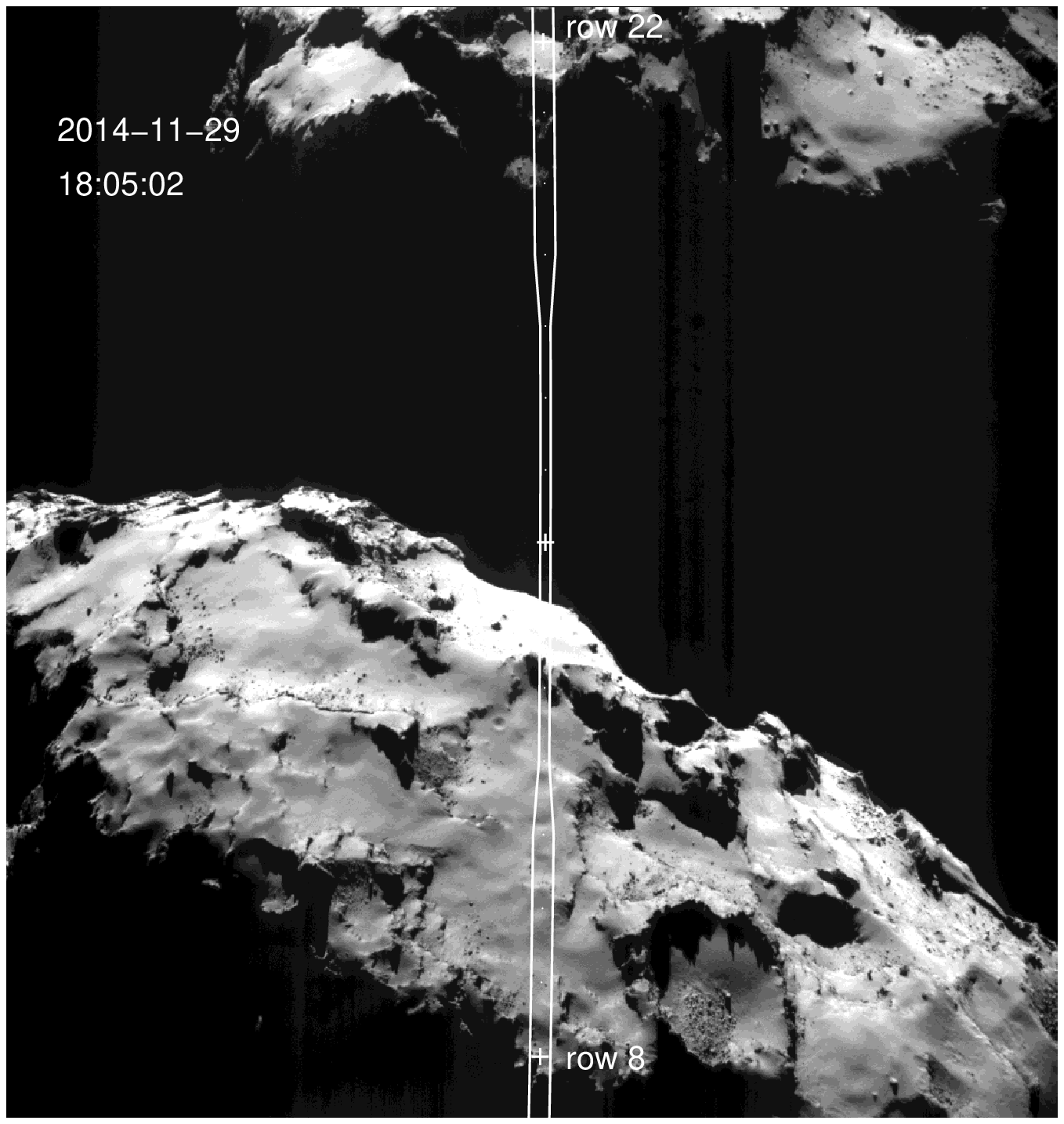}
\includegraphics*[width=0.62\textwidth,angle=0.]{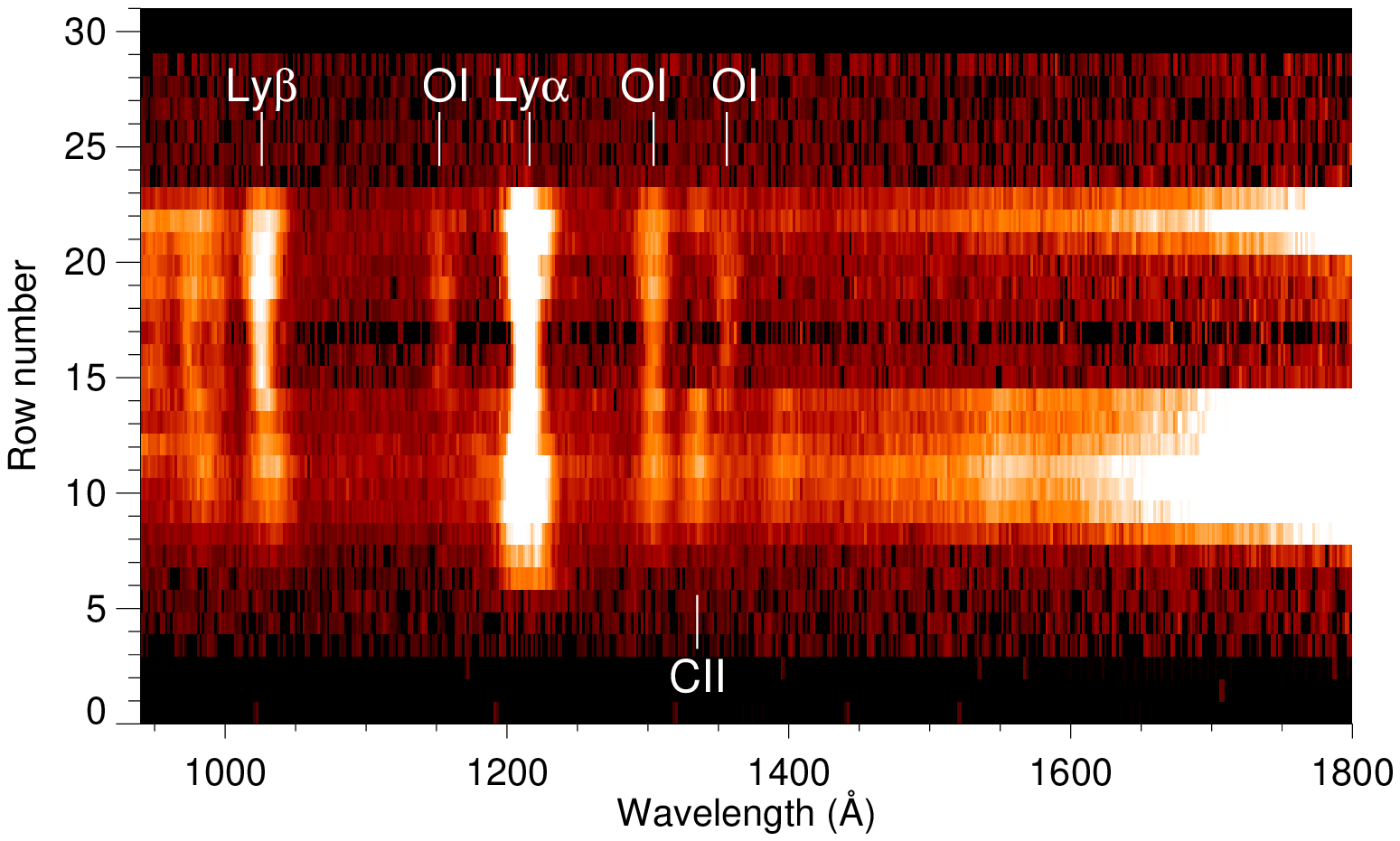}
\caption[]{Left: NAVCAM context image obtained November 29 UT 18:05.  Right: Near simultaneous spectral image beginning UT 18:00, 2419 s exposure, viewing against the dark neck.  The sunlit nucleus appears in rows 7--14 and 20--22.  Note the extension of the coma emissions into rows 20--22.  The distance to the comet was 30.3 km, the heliocentric distance was 2.88~AU, and the solar phase angle was 93.1\degr. \label{nav1} }
\end{center}
\end{figure*}

\begin{figure*}[ht]
\begin{center}
\includegraphics*[width=0.37\textwidth,angle=0.]{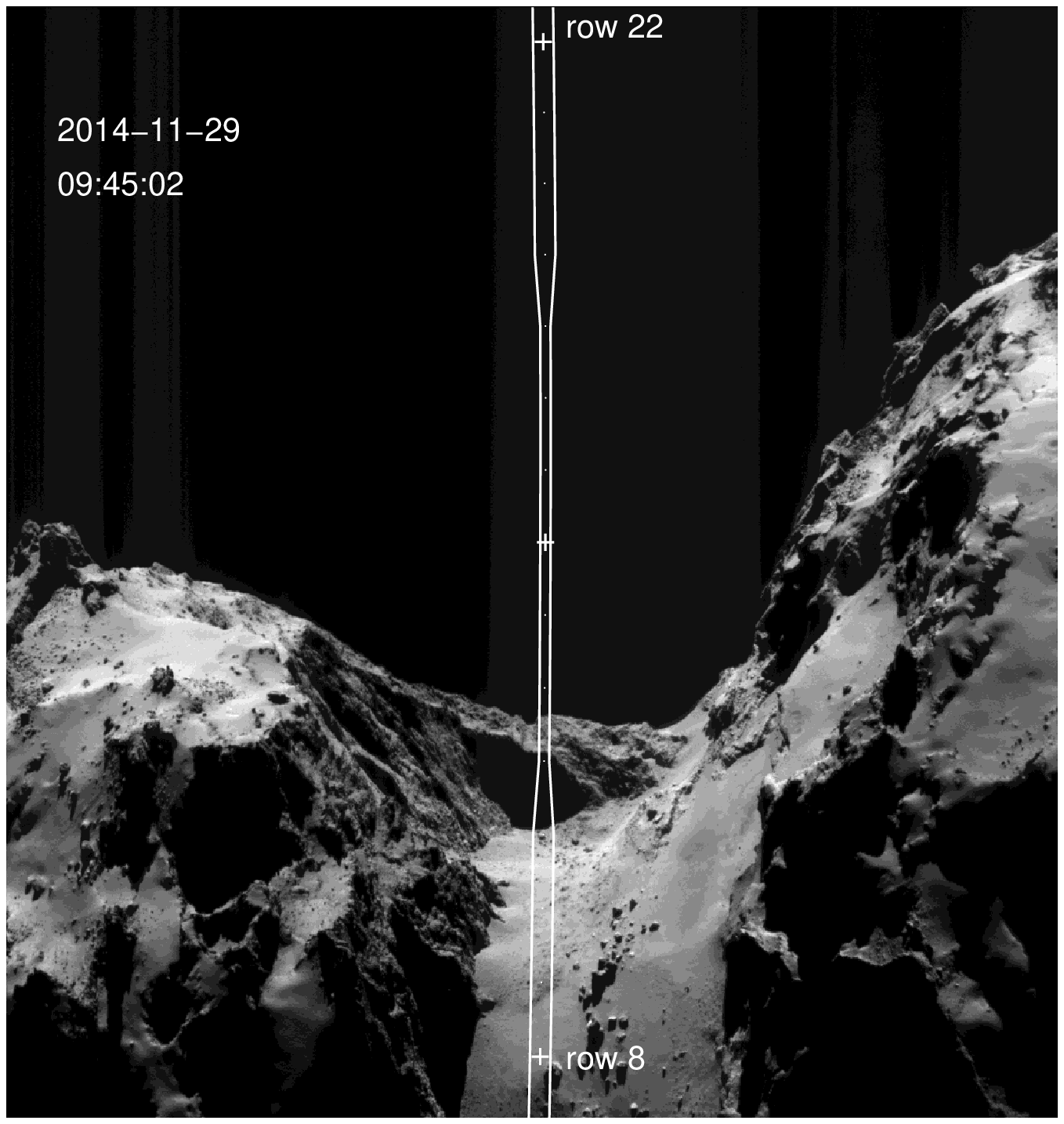}
\includegraphics*[width=0.62\textwidth,angle=0.]{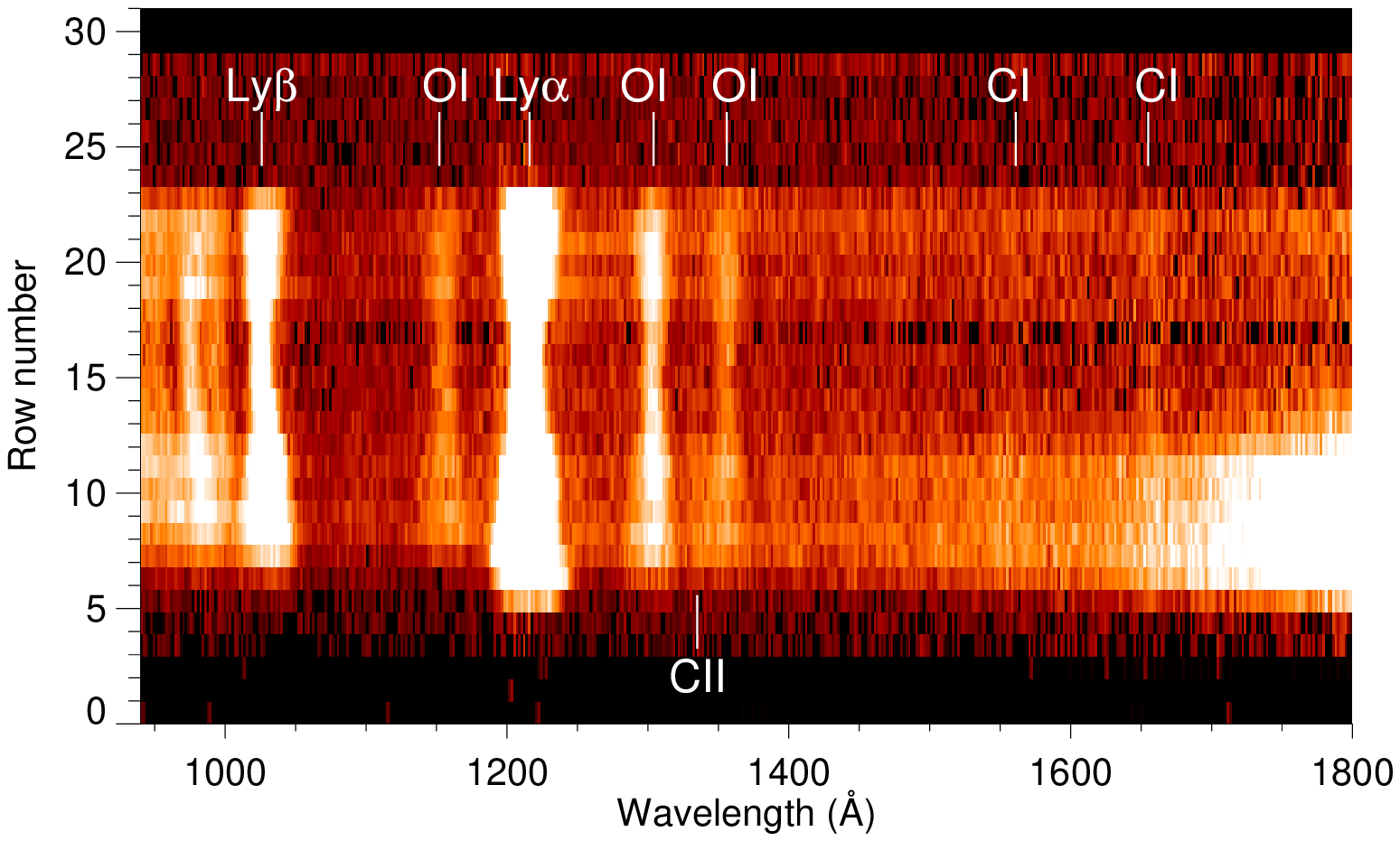}
\caption[]{Left: NAVCAM context image obtained November 29 UT 09:45.  Right: Near simultaneous spectral image beginning UT 09:48, 1209 s exposure.  The coma emissions are visible (including \ion{C}{i}) along the entire length of the slit including the sunlit neck (rows 7--12).  The distance to the comet was 30.3 km and the solar phase angle was 93.6\degr.  \label{nav2} }
\end{center}
\end{figure*}

\begin{figure*}[ht]
\begin{center}
\includegraphics*[width=0.48\textwidth,angle=0.]{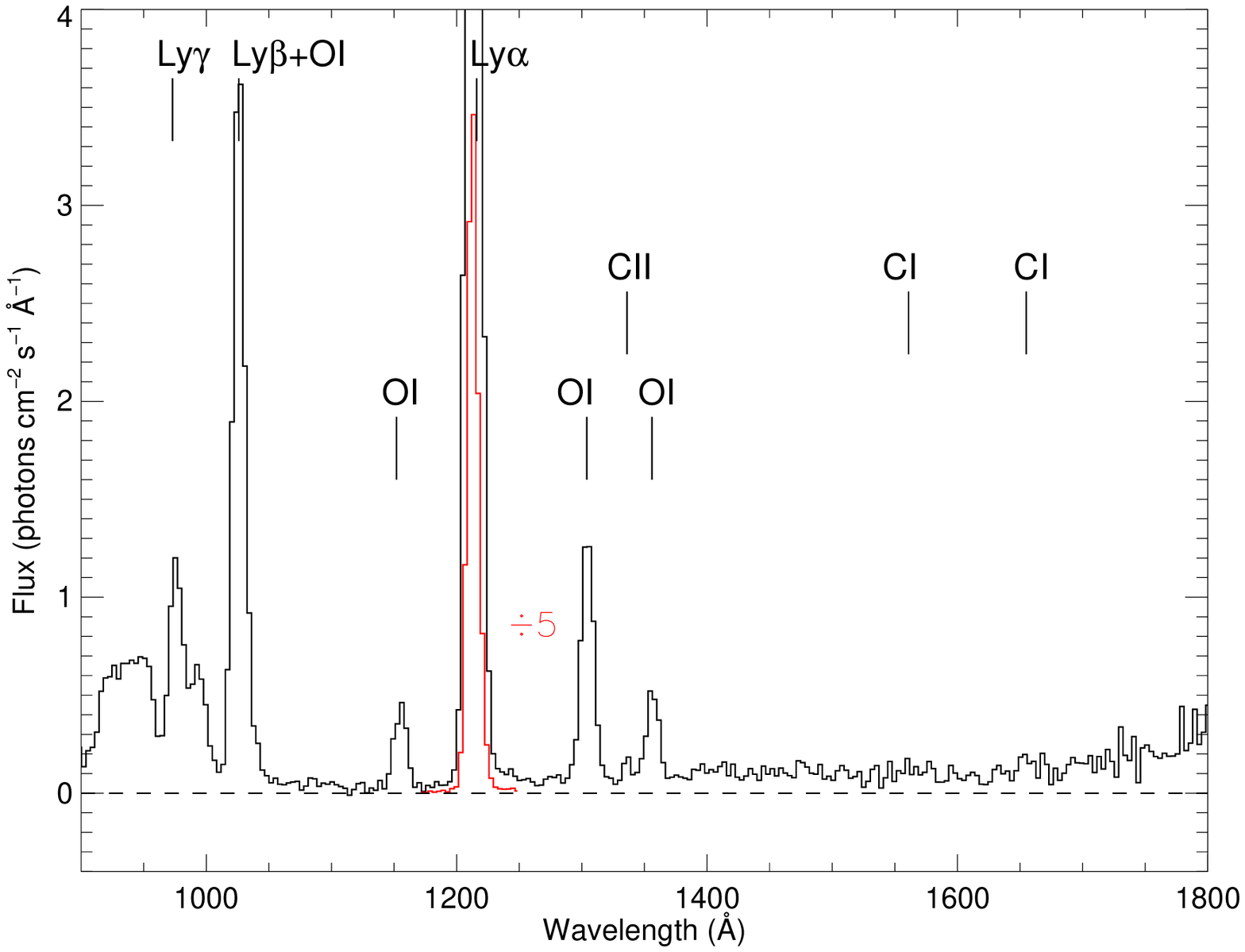}
\includegraphics*[width=0.48\textwidth,angle=0.]{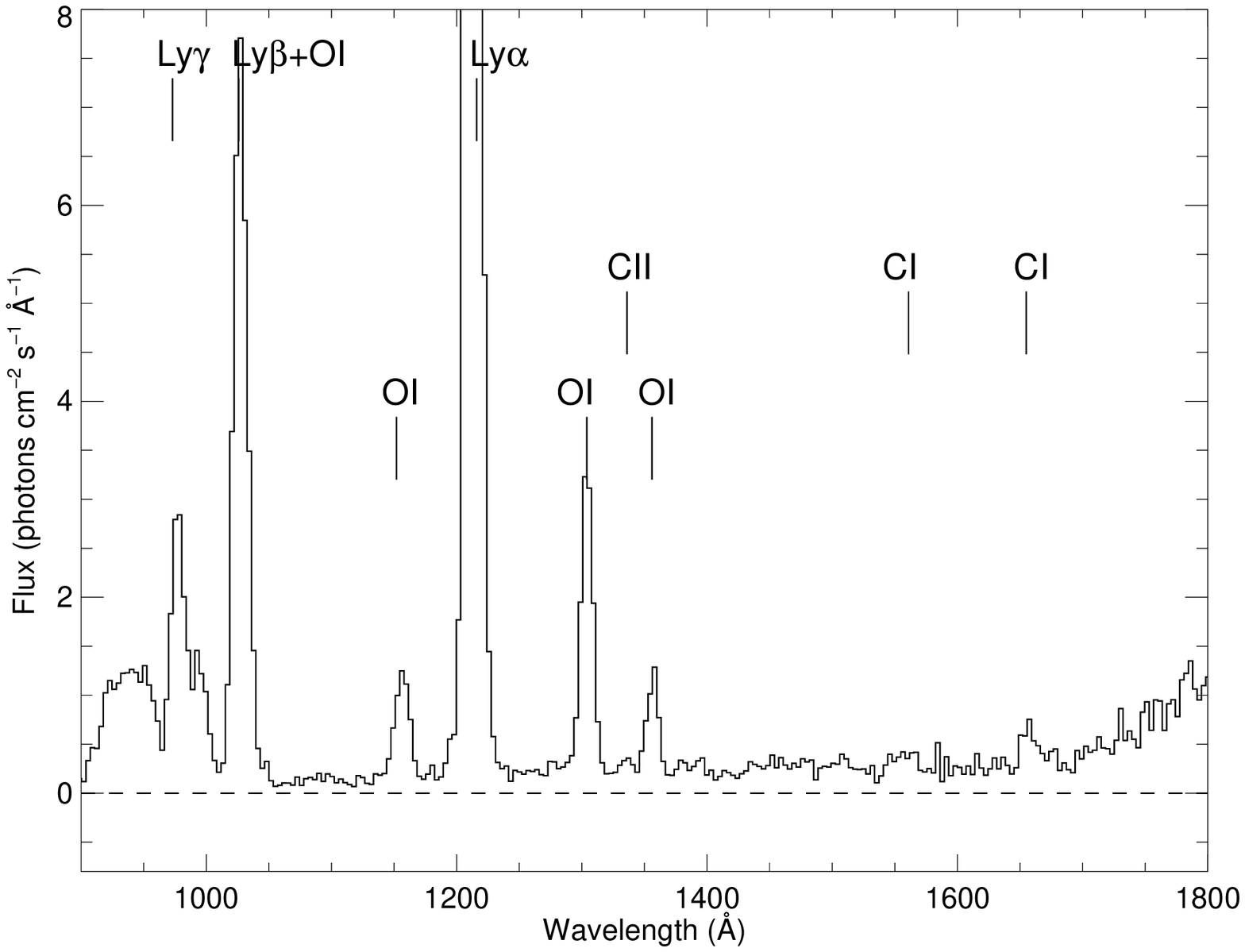}
\caption[]{Spectra obtained from the narrow center of the Alice slit corresponding to the spectral images of Fig.~\ref{nav1}  (left) and Fig.~\ref{nav2} (right).  \label{novspec} }
\end{center}
\end{figure*}

	Following Lander delivery on November 12, Rosetta was placed in a near-circular ($\sim$30~km) near-terminator orbit that permitted Alice to observe \ion{H}{i} and \ion{O}{i} emissions from the coma along a line-of-sight to un-illuminated regions of the nucleus.  One particular example, from November 29, is shown in Figure~\ref{nav1}, consisting of a NAVCAM context image and the corresponding spectral image.  Another example, from the same day, shown in Figure~\ref{nav2}, shows the Alice slit crossing the limb above the neck region and extending into the jet above the neck.  In the first example, the spectrum from the narrow center of the slit, shown in the left panel of Figure~\ref{novspec}, was obtained viewing against the dark neck.  In this case, there is no interplanetary hydrogen background and the ratio of Lyman-$\beta$ to Lyman-$\alpha$ can be used to test our interpretation of electron impact dissociation of \htwoo.  The caveat here is that due to the nature of the detector photocathode \citep{Stern:2007}, the Lyman-$\alpha$ sensitivity varies by a factor of 2 along the slit, as noted from measurements of the interplanetary H emissions made during the Rosetta fly-by of Mars in 2007 \citep{Feldman:2011}.  From the figure, in this spectrum the Lyman-$\alpha$/Lyman-$\beta$ ratio is $\approx$5, whereas the experimental value, for 100 eV electrons, is 7 \citep{Makarov:2004}. For optically thin resonance scattering of solar radiation the Lyman-$\alpha$/Lyman-$\beta$ ratio is $\sim$300, similar to the interplanetary background value.  Considering the above caveats plus the calibration uncertainty at  Lyman-$\alpha$ due to charge depletion of the microchannel plate detector, the low value found for the Lyman-$\alpha$/Lyman-$\beta$ ratio provides confirmation of electron dissociative excitation as the source of the observed atomic emissions.  Note the weak \ion{C}{i} emissions indicating a low relative abundance of \cotwo\ to \htwoo\ in this spectrum.

		The viewing geometry of Figure~\ref{nav2} permitted Alice measurements of the coma emissions above the neck at high northern latitude.  Here the coma emissions are seen both above and against the neck.  The \ion{C}{i} emissions also appear weakly across the entire length of the Alice slit.  The emissions observed above the neck (right panel of Fig.~\ref{novspec}) are brighter than those looking directly at the neck as a result of the Alice line-of-sight extending through the day side of the coma.  

\section{Discussion}

	The far-ultraviolet spectra of the coma of 67P/Churyumov-Gerasimenko as observed by Alice consist predominantly of \ion{H}{i} and \ion{O}{i} emissions and the relative intensities of the strongest of these emissions have not changed over 4 months of observation as the comet approached the Sun from a heliocentric distance of 3.6~AU to 2.8~AU.  These spectra, obtained at high spatial resolution close to the nucleus, are fundamentally different from far-ultraviolet comet spectra observed from Earth orbit, which view the coma on scales of hundreds to thousands of km, in which atomic emissions are primarily due to resonance fluorescence of solar UV radiation by dissociation products of the primary molecular constituents of the nucleus.

Assuming the observed H and O emissions are primarily produced by electron impact dissociation of \htwoo, these far-ultraviolet observations could be mapping the spatial distribution of water plumes erupting from the surface of the nucleus. A similar interpretation has been applied to the mapping of H and O emissions from water plumes emanating from Jupiter's satellite Europa \citep{Roth:2014}. We follow the approach of \citet{Roth:2014} to estimate the electron density required to produce the observed emission brightness, $B_{\lambda}$, at a given wavelength, $\lambda$, in rayleighs (1 rayleigh $= 10^6/4\pi$~photons cm$^{-2}$ s$^{-1}$ sr$^{-1}$).
\[ B_{\lambda} = 10^{-6} \int n_e(z) n_{\rm H_2O}(z) g_{\lambda}(T_e) dz \]
\noindent where $n_e$ and $n_{\rm H_2O}$ are the local electron and \htwoo\ densities, respectively, $g_{\lambda}$ is the electron excitation rate as a function of electron temperature, $T_e$, and the integral is along the line-of-sight to the coma.  If we replace $n_e(z)$ with a mean electron density, ${\bar{n}_e}$, in the interaction region, the equation simplifies to
\[ B_{\lambda} = 10^{-6} \bar{n}_e g_{\lambda}(T_e) \bar{N}_{\rm H_2O}   \]
\noindent where $\bar{N}_{\rm H_2O}$ is the water column density along the line-of-sight.  

To evaluate the excitation rate, $g_{\lambda}$, we approximate the photoelectron energy spectrum with a Maxwellian distribution characterized by an electron temperature of 25 eV, neglecting the low energy component of the energy distribution \citep{Koros:1987} as the dissociative cross sections decrease rapidly below this energy \citep{Makarov:2004,Itikawa:2005}.  We consider the line-of-sight to the nucleus shown in Fig.~\ref{nav1} and its corresponding spectrum in the left panel of Fig.~\ref{novspec}.  For the \ion{H}{i} \lyb\ emission, the observed flux translates to $\sim$30 rayleighs, and the calculated emission rate is $2 \times 10^{-10}$ cm$^3$~s$^{-1}$.  Thus, for a water column density of $1 \times 10^{15}$~cm$^{-2}$ (the low end of the range reported by MIRO, \citet{Gulkis:2015}) the required mean electron density would be $\sim$100 cm$^{-3}$.  This order of magnitude estimate is consistent with both modeled cometary photoelectron fluxes \citep{Koros:1987} and early measurements reported by the IES instrument on Rosetta (J. L. Burch, private communication).  We note that since the photoelectron production rate is also proportional to the \htwoo\ density, the atomic emissions observed by Alice should vary more strongly with position in the coma than the measured molecular column densities. 

In addition to the atomic emissions, spectra taken at lower phase angles show the presence of dust through scattering of long wavelength solar ultraviolet photons with similar spatial dependence as the atomic emissions.  Surprisingly, CO Fourth Positive band emission, routinely observed in Earth-orbital observations of comets, is clearly observed only occasionally and at relatively low brightness, as in the right panel of Figure~\ref{sep_spec}.  Both spectra in this figure show weak \ion{C}{i} multiplets whose relative intensities suggest electron impact dissociation of \cotwo\ \citep{Ajello:1971a,Mumma:1972}.  However, electron impact dissociation of \cotwo\ has a considerably smaller cross section for the production of the CO bands than for \ion{C}{i} $\lambda$1657 \citep{Mumma:1971} and so the absence of CO emission in the left panel suggests that the observed emission in the right panel is due to solar resonance fluorescence.  From the model of \citet{Lupu:2007} the observed band brightnesses correspond to CO line-of-sight column densities $\leq 1 \times 10^{14}$~cm$^{-2}$.  We note that solar resonance fluorescence does not contribute significantly to any of the observed atomic emissions because of the very low column densities of these dissociation products near the nucleus.

\section{Conclusion}

In this initial report we have presented a few examples of the far-ultraviolet spectra obtained by Alice through the end of November 2014.  Limb emission is found to be highly variable with position and time, and during late 2014 appears strongest at northern sub-spacecraft latitudes, and in the vicinity of the comet's ``neck''.  Relative intensities of the strongest \ion{H}{i} and \ion{O}{i} emissions have not varied significantly during this time period and support the interpretation of electron dissociative excitation of \htwoo\ as the primary source of the emission.  \ion{C}{i} emissions, similarly attributed to electron dissociative excitation of \cotwo, do vary relative to  \ion{H}{i} and \ion{O}{i}, consistent with heterogeneous coma composition reported by other Rosetta instruments.  Although the spectrograph slit is not perfectly aligned with the visible jets in our observations, the spatial variation of the emissions along the slit indicates that the excitation occurs within a few hundred meters of the surface and the gas and dust production are correlated.  More detailed future analyses, combining the Alice spectra with data obtained by the other remote sensing instruments and in situ particle measurements, are forthcoming.

\begin{acknowledgements}

	Rosetta is an ESA mission with contributions from its member states and NASA.  We thank the members of the Rosetta Science Ground System and Mission Operations Center teams, in particular Richard Moissl and Michael K\"uppers, for their expert and dedicated help in planning and executing the Alice observations.  We thank Darrell Strobel and Michael Mumma for helpful discussions regarding the electron dissociative excitation of \htwoo\ and \cotwo.  The Alice team acknowledges continuing support from NASA's Jet Propulsion Laboratory through contract 1336850 to the Southwest Research Institute.  The work at Johns Hopkins University was supported by a subcontract from Southwest Research Institute.

\end{acknowledgements}


\clearpage

\end{document}